\documentclass{article}
\usepackage[preprint]{spconfa4}
\usepackage{spconfa4,amsmath,amssymb,graphicx,times}
\usepackage{color}
\usepackage{flushend}
\usepackage{gensymb}
\usepackage[sort]{cite}
\usepackage{subcaption}
\DeclareMathOperator{\EX}{\mathbb{E}}
\copyrightnotice{978-1-5386-8151-0/18/\$31.00~\copyright2018 IEEE}

\title{ON THE DIFFERENCE-TO-SUM POWER RATIO OF SPEECH AND WIND NOISE BASED ON THE CORCOS MODEL}

\twoauthors{Daniele Mirabilii}{}{Emanu\"{e}l A.P. Habets}{} 
\address{International Audio Laboratories Erlangen*,  Am Wolfsmantel 33, 91058 Erlangen, Germany\\
	\{daniele.mirabilii,emanuel.habets\}@audiolabs-erlangen.de \thanks{*A joint institution of the Friedrich-Alexander-University Erlangen-N\"urnberg (FAU) and Fraunhofer IIS, Germany.}}

\begin{document}

\maketitle

\begin{abstract}
The difference-to-sum power ratio was proposed and used to suppress wind noise under specific acoustic conditions. In this contribution, a general formulation of the difference-to-sum power ratio associated with a mixture of speech and wind noise is proposed and analyzed. In particular, it is assumed that the complex coherence of convective turbulence can be modelled by the Corcos model. In contrast to the work in which the power ratio was first presented, the employed Corcos model holds for every possible air stream direction and takes into account the lateral coherence decay rate. The obtained expression is subsequently validated with real data for a dual microphone set-up. Finally, the difference-to-sum power ratio is exploited as a spatial feature to indicate the frame-wise presence of wind noise, obtaining improved detection performance when compared to an existing multi-channel wind noise detection approach.
\end{abstract}


\section{Introduction}

Wind noise often impairs the quality and the intelligibility of speech signals recorded in outdoor environments. The majority of wind noise estimation and reduction approaches developed in the past decade \cite{hofmann2012morphological, nemer2013single, Nelke2014, nelke2015wind } rely on a detection phase to determine the wind noise presence within a short-time frame. The benefit of detecting wind noise before applying a reduction is twofold. Firstly, the detection results can be exploited to further improve the noise estimation and reduction. Secondly, in case of noise absence within a specific time frame, it is possible to reduce the computational complexity of the noise reduction algorithm by leaving the received signal unprocessed.  

Various single-channel detection methods have been developed \cite{ hofmann2012morphological, nemer2013single, nelke2016wind,nelke2014dual}. State-of-the-art device architectures (e.g., smartphones or hearing aids) are often equipped with more than one microphone and therefore a multi-channel processing can be performed, as in \cite{thuene2016maximum,park2016coherence,franz2010multi,elko2016noise}. In particular, existing multi-channel algorithms are based on either time or frequency domain-based detection approaches \cite{zakis2014robust,petersen2008device}. 

In \cite{elko2016noise}, the authors employ a two-element adaptive differential microphone, and apply a multi-channel wind noise suppression based on a frequency-dependent quantity, i.e., the ratio of the difference-signal power to the sum-signal power. In the following, the latter is denoted as difference-to-sum power ratio or simply power ratio. The main idea presented in \cite{elko2016noise} is to exploit the difference between the speed of propagation of acoustic sources (e.g., speech) and the speed of propagation of convective turbulence (e.g., a wind flow). Assuming equivalent power levels at the microphones, this results in a strong dissimilarity between the output signal ratio of speech and turbulent signals, since the speed of propagation of air flows is much lower than the speed of propagation of radiating acoustic sources. Therefore, the power ratio associated to wind noise presents higher values with respect to the power ratio associated to speech, resulting in a reliable feature to detect which frequency band is more likely to be distorted by wind-induced noise.  

In \cite{elko2016noise,elko2007reducing}, it is assumed that the complex coherence of wind noise can be predicted by the Corcos model, originally presented in \cite{corcos1964structure}, which describes the stochastic pressure distribution of convective turbulence in a turbulent boundary layer. Following the Corcos model, the pressure field of an air stream loses coherence with frequency-dependent longitudinal (parallel to the stream direction) and lateral (orthogonal to the stream direction) exponential decays depending on the free-field air velocity and the displacement from which the field is observed, showing higher magnitude coherence values for very low frequency ranges. However, the authors in \cite{elko2016noise} assume an isotropic coherence decay which is dependent only on the microphone distance and the free-field air velocity: the dependency on the stream direction and the lateral decay is not considered.
 
In this contribution, a definition of the Corcos model is used that holds for every possible direction of the air stream and includes the lateral coherence decay, as shown in \cite{mirabilii2018simulating}. In addition, we propose a different definition for the additive model of the signals captured by the microphones. In particular, we assume two different wind noise contributions which exhibit a complex coherence given by the Corcos model, rather than assuming a wind noise contribution at the reference microphone and the same contribution with a phase difference at the second microphone as in \cite{elko2016noise}. We obtain a final expression of the power ratio of pure wind noise which differs from the one from \cite{elko2016noise}, and validate the correctness of this expression for the power ratio of clean speech. Simulation results on the power ratio of real wind noise recordings show a close match with the theoretical definition formulated in this contribution. Finally, we develop a wind noise detector based on the proposed expression of the power ratio, obtaining an improved accuracy with respect to the magnitude squared coherence-based detector presented in \cite{NelkePhD2016}.
  
\section{Signal model}

Let us define the observed signals as a superposition of anechoic speech and wind noise. Given a microphone distance denoted by $d$, the signal model in the discrete Fourier transform (DFT) domain can be expressed as
\begin{equation}
X_1(\omega_k) = S(\omega_k) +V_1(\omega_k) \label{eq:1}
\end{equation} 
\begin{equation}
X_2(\omega_k) = S(\omega_k)\cdot e^{-j\omega_k\tau_s} + V_2(\omega_k) \label{eq:2}
\end{equation} where the subscripts denote the index of the corresponding microphone, $\omega_k$ denotes the discrete angular frequency, $S(\omega_k)$ denotes the speech signal, $V_1(\omega_k)$ and $V_2(\omega_k)$  denote the two wind noise contributions and $\tau_s = d \cos(\theta_\textrm{s})/c$ denotes the time difference of arrival (TDOA), with $\theta_\textrm{s}$ denoting the direction of arrival (DOA) of the speech with respect to the microphone axis and $c$ denoting the speed of propagation of radiating acoustic sources expressed in $\textrm{ms}^{-1}$. We assume the speech and the wind noise contributions to be uncorrelated. The wind noise contributions $V_1(\omega_k)$ and $V_2(\omega_k)$ are assumed to exhibit a complex coherence approximated by the Corcos model as in \cite{mirabilii2018simulating}, given by
\begin{equation}
\gamma_c(\omega_k) = \exp{\left ( \frac{-\alpha(\theta_\textrm{w}) \omega_k d}{ U_c }  \right )} \exp{\left ( \frac{\iota \: \omega_k d\cos(\theta_\textrm{w})}{ U_c }  \right )}, \label{eq:3} 	
\end{equation} where $\iota = \sqrt{-1}$, $\theta_\textrm{w}$ denotes the DOA of the wind stream with respect to the microphone axis, $\alpha(\theta_\textrm{w})$ denotes a DOA-dependent decay rate parameter which is defined as
\begin{equation}
\alpha(\theta_\textrm{w})=\alpha_1\mid\cos(\theta_\textrm{w})\mid+\alpha_2\mid\sin(\theta_\textrm{w})\mid,\label{eq:4}
\end{equation} where $\alpha_1$ and $\alpha_2$ are the longitudinal and the lateral coherence decay rates respectively,  experimentally determined in \cite{mellen1990modeling}. Finally, $U_c$ is the convective turbulence speed in a boundary layer, where $U_c \approx 0.8  U$, with $U$ denoting the free-field wind stream velocity. We can further assume spatially white wind noise contributions for a sufficiently large microphone distance $d$ by modifying the final expression of the power ratio with $\gamma_c(\omega_k) = 0$. 
\section{Difference-to-sum Power Ratio}
Given the difference signal in the DFT domain
\begin{equation}
X_{\mathrm{diff}}(\omega_k) = X_1(\omega_k) - X_2(\omega_k),  \label{eq:5}
\end{equation} we compute the power spectral density (PSD) defined by
\begin{equation}
\Phi_{\mathrm{diff}}(\omega_k) = \EX{ \{ X_{\mathrm{diff}}(\omega_k) X^*_{\mathrm{diff}}(\omega_k) \} },\label{eq:6}
\end{equation} where $\EX{ \{ . \} }$ denotes the expected value. Assuming equal power for the wind noise contributions at each microphone, i.e.,
\begin{equation}
\Phi_{V_1V_1}(\omega_k) = \Phi_{V_2V_2}(\omega_k) =\Phi_{vv}(\omega_k) \label{eq:7}
\end{equation} and exploiting trigonometric equivalences we obtain 
\begin{equation}
\Phi_{\mathrm{diff}} =  4\Phi_{ss}\sin^2\left ( \frac{\omega_k d_{\theta_\textrm{s}}}{2c} \right ) + 2\Phi_{vv} \left [ 1 - \operatorname{Re}\{\gamma_c\} \right ], \label{eq:A8}
\end{equation} where the dependency on $\omega_k$ is omitted for brevity, $\Phi_{ss}(\omega_k)$ denotes the speech signal PSD, $\Phi_{vv}(\omega_k)$ denotes the wind noise PSD, $ d_{\theta_s} = d \cos(\theta_\textrm{s})$ and $\operatorname{Re}\{.\}$ denotes the real part operator. Likewise, let us define the sum signal
\begin{equation}
X_{\mathrm{sum}}(\omega_k) = X_1(\omega_k) + X_2(\omega_k),  \label{eq:9}
\end{equation} and compute the PSD
\begin{equation}
\Phi_{\mathrm{sum}}(\omega_k) = \EX{ \{ X_{\mathrm{sum}}(\omega_k) X^*_{\mathrm{sum}}(\omega_k) \} }.\label{eq:10}
\end{equation}Using the same procedure as with the difference PSD we obtain
\begin{equation}
\Phi_{\mathrm{sum}} =  4\Phi_{ss}\cos^2\left ( \frac{\omega_k d_{\theta_s}}{2c} \right ) + 2\Phi_{vv} \left [ 1 + \operatorname{Re}\{\gamma_c\} \right ]. \label{eq:11}
\end{equation} Given the definition of the power ratio 
\begin{equation}
\textrm{PR}(\omega_k)=\frac{\Phi_{\mathrm{diff}}(\omega_k)}{\Phi_{\mathrm{sum}}(\omega_k)} \label{eq:12}
\end{equation} and exploiting (\ref{eq:9}) and (\ref{eq:11}) we finally obtain
\begin{equation}
\textrm{PR}(\omega_k)=\frac{ 4\Phi_{ss}\sin^2\left ( \frac{\omega_k d_{\theta_\textrm{s}}}{2c} \right ) + 2\Phi_{vv} \left [ 1 - \operatorname{Re}\{\gamma_c\} \right ]}{4\Phi_{ss}\cos^2\left ( \frac{\omega_k d_{\theta_\textrm{s}}}{2c} \right ) + 2\Phi_{vv} \left [ 1 + \operatorname{Re}\{\gamma_c\} \right ]}.\label{eq:13}
\end{equation} It is possible to define the clean speech signal power ratio by setting the wind noise PSD to zero ($\Phi_{vv}(\omega) = 0$), obtaining
\begin{equation}
\textrm{PR}_\textrm{s}(\omega_k) = \tan^2\left ( \frac{\omega_k d \cos(\theta_\textrm{s})}{2c} \right ),\label{eq14}
\end{equation} which presents the same expression as in 	\cite{elko2016noise}. Increasing the spacing $d$ results in a periodic behaviour of the power ratio of the clean speech signals for $\theta_\textrm{s} \neq 90 \degree$, leading to more asymptotes in the frequency range and therefore higher values. For pure wind noise ($\Phi_{ss}(\omega) = 0$), the power ratio is given by
\begin{equation}
\textrm{PR}_\textrm{w}(\omega_k)=\frac{1 - \exp \left ({\frac{-\alpha(\theta_\textrm{w}) \omega_k  d}{U_c}} \right )  \cos \left ( \frac{ \omega_k  d \cos(\theta_\textrm{w}) }{U_c}  \right )   }{1 + \exp \left ({\frac{-\alpha(\theta_\textrm{w})  \omega_k  d}{U_c}} \right )  \cos \left ( \frac{ \omega_k d  \cos(\theta_\textrm{w}) }{U_c}  \right )}.\label{eq15}
\end{equation} 
\begin{figure}[!t]
	\centering
	\includegraphics[width=1\linewidth]{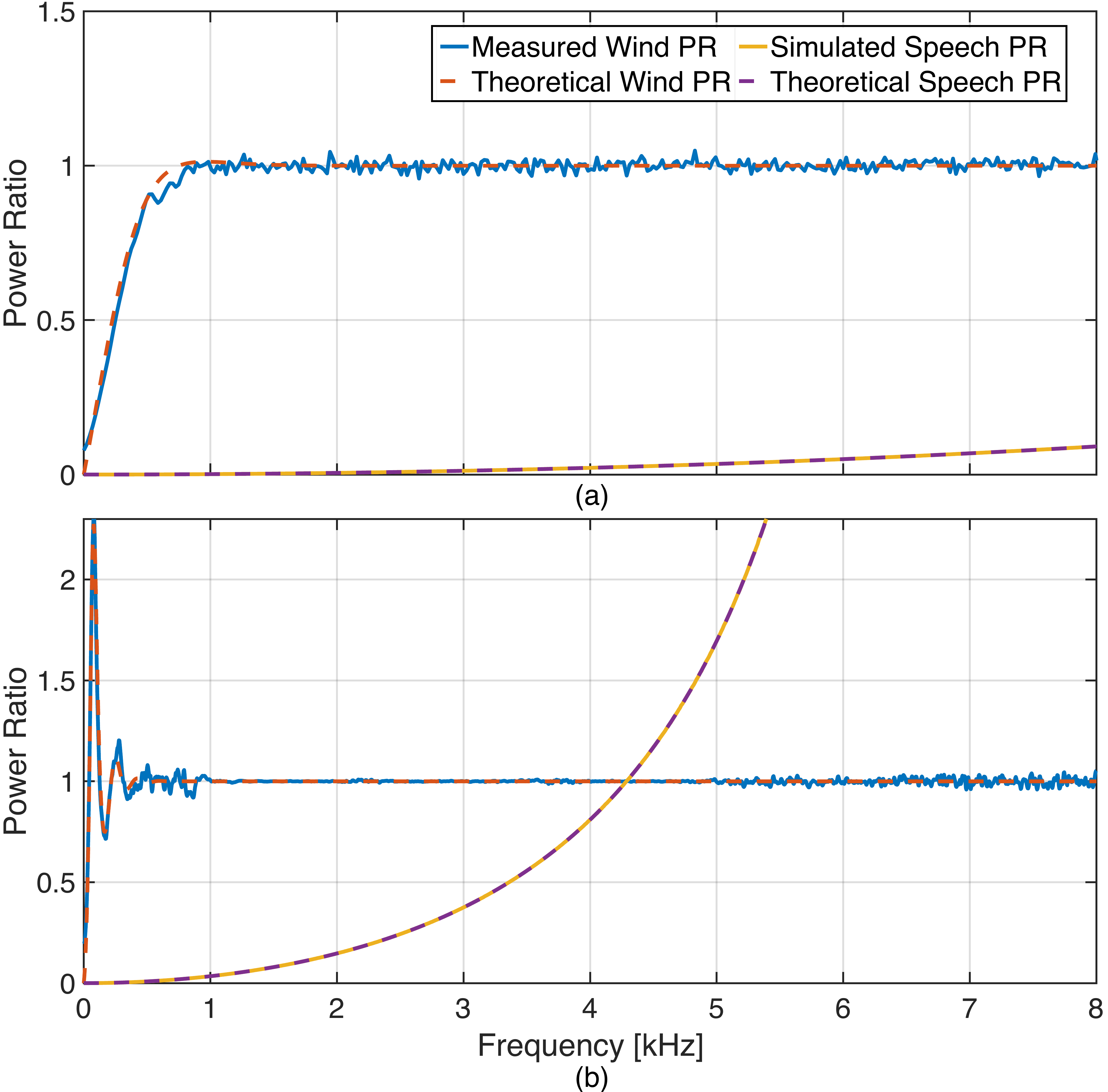}
	\caption{Wind noise power ratio compared to the speech power ratio with  (a) $d = 4$ mm, $\theta_\textrm{w} = 90 \degree$, $U = 1.8 $ m/s and $\theta_\textrm{s} = 0 \degree$ (b) d = 20 mm, $\theta_\textrm{w} = 0 \degree$, $U = 2.8 $ m/s and $\theta_\textrm{s} = 0 \degree$. } \label{fig:1}
\end{figure}
From (\ref{eq15}) it is clear that for larger distances $d$ the exponential decay of the complex coherence of the wind noise $\gamma_c(\omega_k)$ rapidly tends toward zero for increasing frequency, so that the wind noise power ratio can be assumed as unitary and frequency independent. Figure \ref{fig:1} shows the comparison between the power ratio of wind noise and the power ratio of clean speech for two different conditions: (a) with a microphone distance of 4 mm, a valuable separation between wind noise and speech is achieved, while for (b) with a microphone distance of 20 mm, the separation is disrupted due to the fact that the speech power ratio increases with frequency for $\theta_\textrm{s} \neq 90 \degree$. Nevertheless, the separation holds in the range 0-1 kHz, where most of the wind noise energy is concentrated. Moreover, (a) depicts the power ratio of wind noise for  $\theta_\textrm{w} = 90 \degree$, characterised by small values in the low-frequency region, increasing towards unity, while (b) depicts the power ratio of wind noise for $\theta_\textrm{w} = 0 \degree$, characterised by oscillations in the low-frequency range. In the latter case the power ratio of wind noise can be greater than one. The measured data of the wind noise power ratio (solid blue lines) closely follows the theoretical expression (dashed red lines) given by (\ref{eq15}). The results shown here were obtained using speech convolved with the direct-path impulse responses and recorded wind noise from the experiment described in \cite{mirabilii2018simulating}. 
\begin{figure}[!t]
	\centering
	\includegraphics[width=1\linewidth]{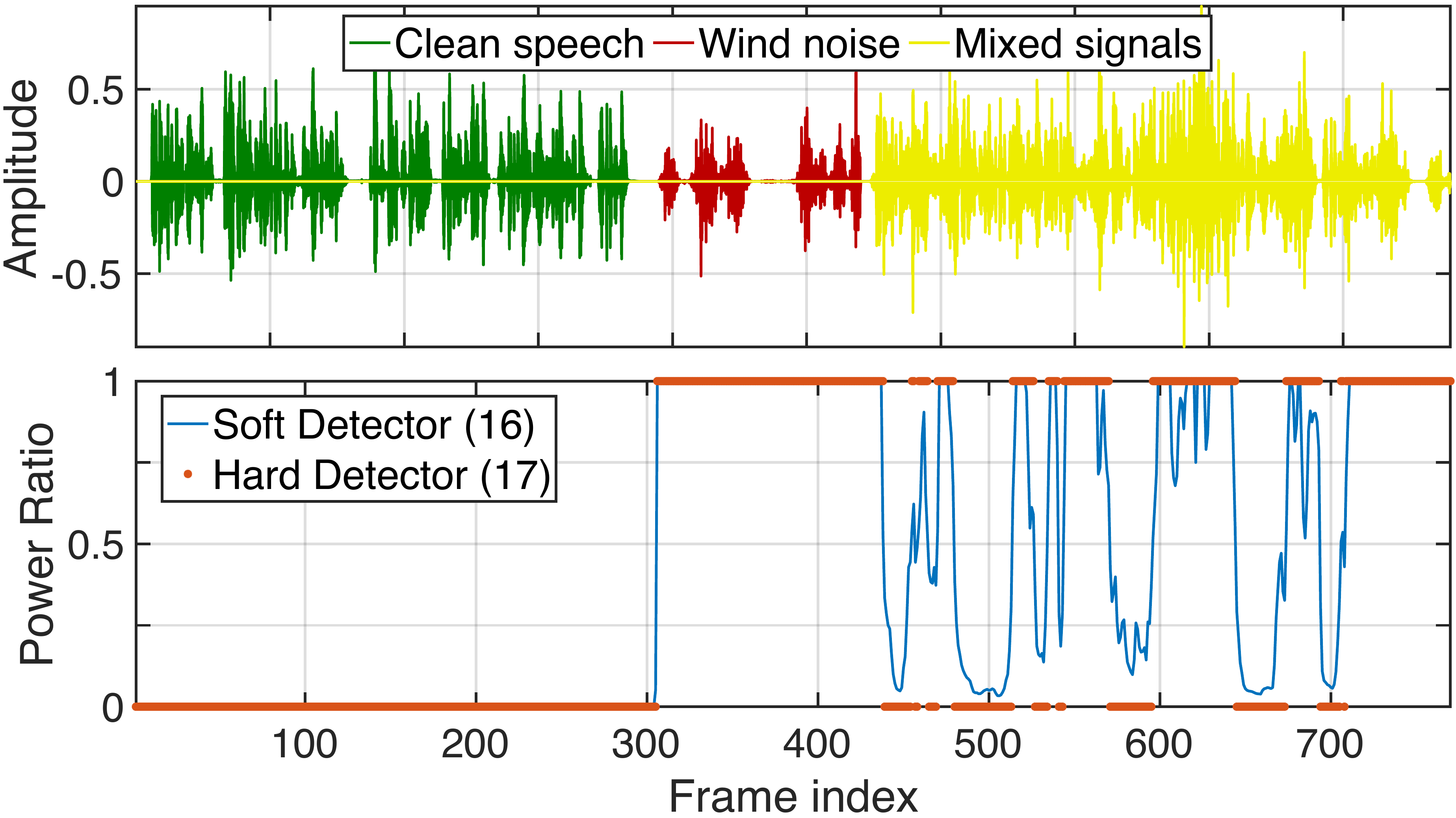}
	\caption{Time signal consecutively containing clean speech ($\theta_\textrm{s} = 90 \degree$), pure wind noise ($\theta_\textrm{w} = 0 \degree$, $U = 1.8$ m/s) and their mixture for $d = 4$ mm. Below, the relative power ratio-based detection outcome, where the hard detector was computed with a threshold $\theta = 0.5$.}	\label{fig:2}
\end{figure}
\section{Wind noise detector}
It is possible to use the expressions in (\ref{eq14}) and (\ref{eq15}) to design a frame-wise wind noise activity detector. By choosing an adequate microphone distance, speech signals present a power ratio that is considerably lower with respect to the wind noise power ratio, so that a reliable separation can be achieved. To obtain a measure to quantify the amount of wind noise distortion within a time frame we define the wind noise soft activity detector as
\begin{equation}
I_{\textrm{PR}}(l) =\frac{1}{\#\mathcal{K}} \cdot \displaystyle \sum_{\omega_k \in \mathcal{K} }^{}\mathrm{PR}(l,\omega_k), \label{eq16}
\end{equation} where $\mathcal{K}$ contains the considered discrete angular frequencies, $\#\mathcal{K}$ denotes the number of frequencies in $ \mathcal{K}$ and $l$ denotes the index of the time frame. The soft detector generally presents values in the range [0,1], where lower values of the power ratio are associated to clean speech. Moreover, it is possible to define a hard detector by thresholding (\ref{eq16}), e.g.,
\begin{equation}
J_{\textrm{PR}}(l) =
\begin{cases}
1 & \quad  \text{for} \quad I_{\textrm{PR}}(l) > \theta \\
0 & \quad  \text{for} \quad I_{\textrm{PR}}(l) \leq \theta \\
\end{cases}, \label{eq17}
\end{equation} where $\theta$ denotes the threshold which can arbitrarily vary in the interval [0,1].  Figure \ref{fig:2} depicts the outcome of the soft and the hard detector applied to a time signal sequentially containing clean speech, pure wind noise and the superposition of the two: the averaged power ratio in (\ref{eq16}) and the hard detector in (\ref{eq17})  categorise speech and wind noise which are associated to lower and higher values of the detectors respectively.  
\section{Performance evaluation}

\begin{figure}[!t]
\centering
\includegraphics[width=1\linewidth]{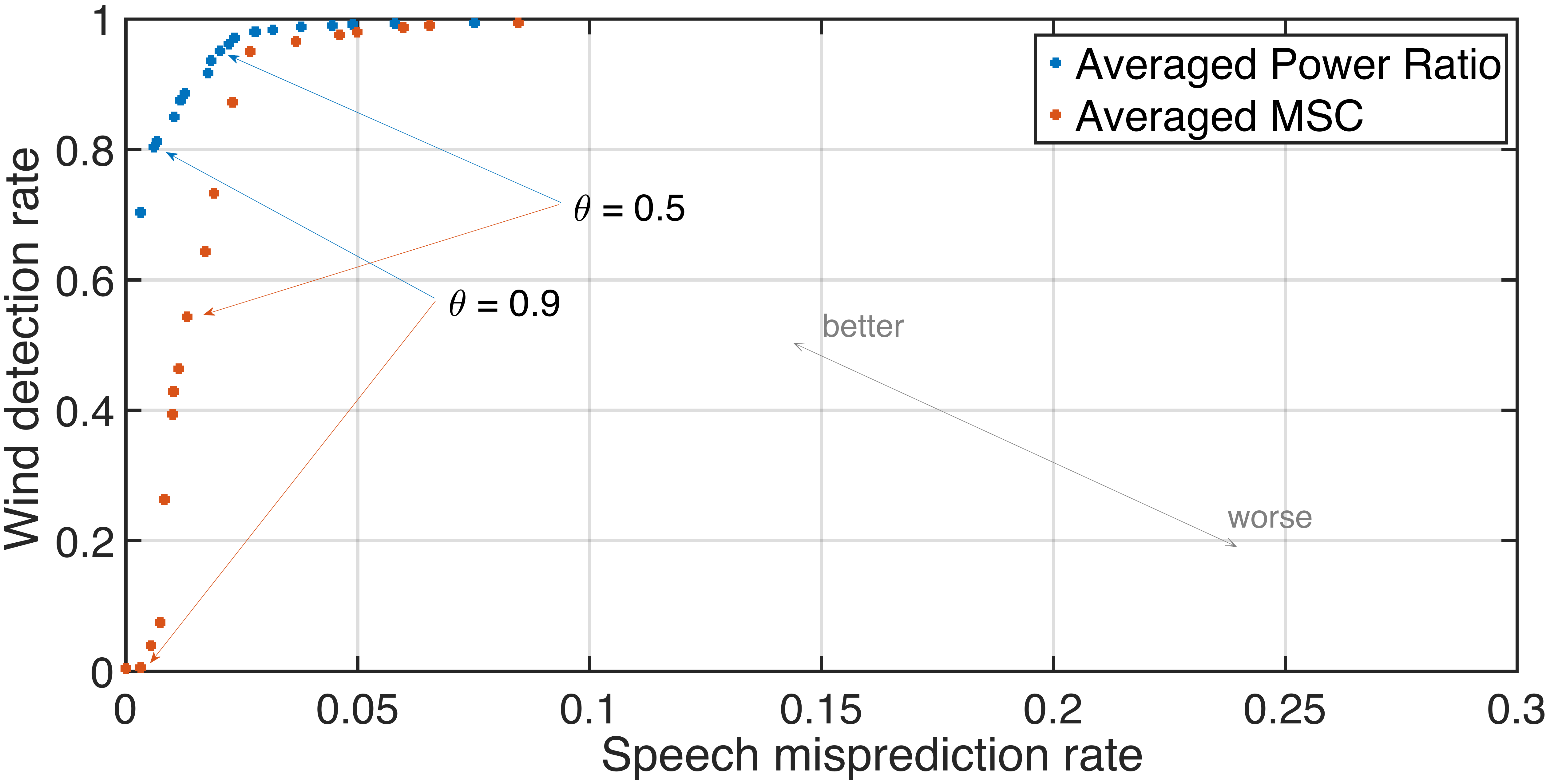}
\caption{ROC of the power ratio-based detector (in blue) and the MSC-based detector (in red).} \label{fig:Figure_8A}
\end{figure}

We assessed the performance of the wind noise detector based on the difference-to-sum power ratio, with an existing multi-channel detection approach, i.e., the magnitude squared coherence-based (MSC) algorithm presented in \cite{NelkePhD2016}, using a receiving operating curve (ROC) comparison. The MSC-based detector is defined as
\begin{equation}
I_{\textrm{MSC}}(l) =1 - \frac{1}{\#\mathcal{K}} \cdot \displaystyle \sum_{\omega_k \in \mathcal{K}}^{}\mathrm{MSC}(l,\omega_k), 
\end{equation} where $\mathrm{MSC}(l,\omega_k)$ indicates the magnitude squared coherence values. The hard decision was computed as in (\ref{eq17}) to obtain $J_{\textrm{MSC}}(l)$. The simulation was carried out using signals sequentially composed of clean speech/pure wind noise/a mixture of clean speech and wind noise at -5 dB of input signal-to-noise ratio (iSNR)/clean speech/a mixture of clean speech and wind noise at -5 dB of iSNR. The alternation of distorted and clean speech was chosen to include onsets/offsets of wind noise. 
The speech and wind noise audio items were randomly selected and subsequently mixed from two different databases: the speech signals were selected from the LibriSpeech ASR Corpus \cite{panayotov2015librispeech} and wind noise signals were selected from a collection of 100 samples of artificially generated dual-microphone wind noise, from the simulation approach presented in \cite{mirabilii2018simulating}. 
The dual-microphone clean speech presents no relative-phase difference, simulating a broad-side scenario. The labelling process was performed manually, associating the label 1 to the frames containing wind noise and speech/wind noise mixture and the label 0 to the frames containing clean speech.  To generate the ROC, we used 20 different values of the hard detection threshold $\theta$ from 0 to 1, with a 0.05 step, for each detector consistently. We finally computed the wind noise detection rate (true positive rate) against the speech misdetection rate (false positive rate). For every chosen value of the threshold $\theta$, we computed the mentioned performance measures, for 10 different and randomly mixed speech and wind noise signals, subsequently averaging the results. Denoting with $L(l)$ the label associated to the $l$-th frame of the processed signal (0 if speech or 1 if wind noise) and with $J(l,\theta)$ the outcome of one of the analysed hard detectors in the $l$-th frame and for the chosen threshold $\theta$, the wind noise detection rate was defined by
\begin{equation}
P_{\textrm{w}}(\theta) = \frac{\sum_{l}^{}Q_\textrm{w}(l,\theta)}{M_\textrm{w}} , 
\end{equation} where $M_\textrm{w}$ denotes the total number of frame labelled as wind noise and
\begin{equation}
Q_\textrm{w}(l,\theta) =
\begin{cases}
1 & \quad  \text{for} \quad J(l,\theta) = 1  \;\text{and}\; L(l) = 1\\
0 & \quad  \text{for} \quad J(l,\theta) = 1  \;\text{and} \; L(l) = 0\\
\end{cases},
\end{equation} while the speech misdetection rate was defined by
\begin{equation}
P_{\textrm{s}}(\theta) = \frac{\sum_{l}^{}Q_\textrm{s}(l,\theta)}{M_\textrm{s}} , 
\end{equation} where $M_\textrm{s}$ denotes the total number of frame labelled as clean speech and
\begin{equation}
Q_\textrm{s}(l,\theta) =
\begin{cases}
1 & \quad  \text{for} \quad J(l,\theta) = 1  \;\text{and}\; L(l) = 0\\
0 & \quad  \text{for} \quad J(l,\theta) = 0  \;\text{and} \; L(l) = 0\\
\end{cases},
\end{equation} for both the analysed hard detectors $J_{\textrm{PR}}(l,\theta)$ and $J_{\textrm{MSC}}(l,\theta)$ consistently. The sampling frequency was 16 kHz and the frame length was 128 ms with 75\% of overlap between consecutive frames. The expected values in (\ref{eq:6}) and (\ref{eq:10}) as well as the ones used to compute the magnitude squared coherence were recursively obtained through a smoothed periodogram, with the smoothing parameter set to 0.5. The computation of both detectors was limited to the range 0-500 Hz. Figure \ref{fig:Figure_8A} shows the ROC for both detectors. It is noticeable how points belonging to the power ratio detector lie on the upper-left section of the ROC with a denser distribution, denoting (a) higher wind detections and lower speech misdetections than the competing detector and (b) less sensitivity to the threshold $\theta$, leading to an improved separation between clean speech and wind noise.
\section{Conclusion}
A general expression of the power ratio for a mixture of speech and wind noise signals was presented. The proposed expression takes into account the air stream direction and the lateral coherence decay rate of the Corcos model for the approximation of the complex coherence of wind noise. Results obtained using measured data showed how closely the power ratio of recorded wind noise matches the proposed theoretical model. Moreover, we exploited the power ratio to design a multi-channel wind noise activity detector, with the aim of identifying the presence of wind noise within each frame. We compared the performance of the proposed detection approach with an existing multi-channel wind noise detector, namely the MSC-based detector. The comparison was based on the receiving operating curves, where the power ratio-based detector outperformed the competing detector in terms of wind detection rate against speech misdetection rate.

\newpage
\pagebreak
\newpage

\bibliographystyle{ieeetr}
\bibliography{ICSSE_2018_PR}

\begin{thebibliography}{10}

\bibitem{hofmann2012morphological}
C.~Hofmann, T.~Wolff, M.~Buck, T.~Haulick, and W.~Kellermann, ``A morphological
  approach to single-channel wind-noise suppression,'' in {\em Proc. Intl.
  Workshop Acoust. Echo Noise Control (IWAENC)}, pp.~1--4, VDE, 2012.

\bibitem{nemer2013single}
E.~Nemer, W.~LeBlanc, M.~Zad-Issa, and J.~Thyssen, ``Single microphone wind
  noise suppression,'' Aug.~20 2013.
\newblock US Patent 8,515,097.

\bibitem{Nelke2014}
C.~M. Nelke, N.~Chatlani, C.~Beaugeant, and P.~Vary, ``Single microphone wind
  noise {PSD} estimation using signal centroids,'' {\em Proc. {IEEE} Intl.
  Conf. on Acoustics, Speech and Signal Processing (ICASSP)}, pp.~7113--7117,
  2014.

\bibitem{nelke2015wind}
C.~M. Nelke and P.~Vary, ``Wind noise short term power spectrum estimation
  using pitch adaptive inverse binary masks,'' in {\em Proc. {IEEE} Intl. Conf.
  on Acoustics, Speech and Signal Processing (ICASSP)}, pp.~5068--5072, 2015.

\bibitem{nelke2016wind}
C.~Nelke, P.~Jax, and P.~Vary, ``Wind noise detection: Signal processing
  concepts for speech communication,'' in {\em DAGA International Conference on
  Acoustics}, 2016.

\bibitem{nelke2014dual}
C.~M. Nelke and P.~Vary, ``Dual microphone wind noise reduction by exploiting
  the complex coherence,'' in {\em Proc. of the {ITG} Conference on Speech
  Communication}, pp.~1--4, VDE, 2014.

\bibitem{thuene2016maximum}
P.~Thuene and G.~Enzner, ``Maximum-likelihood approach to adaptive
  multichannel-wiener postfiltering for wind-noise reduction,'' in {\em Proc.
  of the {ITG} Conference on Speech Communication}, pp.~1--5, VDE, 2016.

\bibitem{park2016coherence}
J.~Park, J.~Park, S.~Lee, J.~Kim, and M.~Hahn, ``Coherence-based dual
  microphone wind noise reduction by wiener filtering,'' in {\em Proceedings of
  the 8th International Conference on Signal Processing Systems}, pp.~170--172,
  ACM, 2016.

\bibitem{franz2010multi}
S.~Franz and J.~Bitzer, ``Multi-channel algorithms for wind noise reduction and
  signal compensation in binaural hearing aids,'' in {\em Proc. Intl. Workshop
  Acoust. Echo Noise Control (IWAENC)}, 2010.

\bibitem{elko2016noise}
G.~W. Elko, J.~M. Meyer, and T.~F. Gaensler, ``Noise-reducing directional
  microphone array,'' Mar.~18 2016.
\newblock US Patent App. 15/073,754.

\bibitem{zakis2014robust}
J.~A. Zakis and C.~M. Tan, ``Robust wind noise detection,'' in {\em Proc.
  {IEEE} Intl. Conf. on Acoustics, Speech and Signal Processing (ICASSP)},
  pp.~3655--3659, 2014.

\bibitem{petersen2008device}
K.~S. Petersen, G.~Bogason, U.~Kjems, and T.~B. Elmedyb, ``Device and method
  for detecting wind noise,'' Mar.~4 2008.
\newblock US Patent 7,340,068.

\bibitem{elko2007reducing}
G.~W. Elko, ``Reducing noise in audio systems,'' Jan.~30 2007.
\newblock US Patent 7,171,008.

\bibitem{corcos1964structure}
G.~Corcos, ``The structure of the turbulent pressure field in boundary-layer
  flows,'' {\em Journal of Fluid Mechanics}, vol.~18, no.~3, pp.~353--378,
  1964.

\bibitem{mirabilii2018simulating}
D.~Mirabilii and E.~A. Habets, ``Simulating multi-channel wind noise based on
  {C}orcos model,'' in {\em Proc. Intl. Workshop Acoust. Echo Noise Control
  (IWAENC)}, 2018.

\bibitem{NelkePhD2016}
C.~M. Nelke, {\em Wind Noise Reduction–Signal Processing Concepts}.
\newblock PhD thesis, RWTH Aachen University, 2016.

\bibitem{mellen1990modeling}
R.~H. Mellen, ``On modeling convective turbulence,'' {\em The journal of the
  Acoustical Society of America}, vol.~88, no.~6, pp.~2891--2893, 1990.

\bibitem{panayotov2015librispeech}
V.~Panayotov, G.~Chen, D.~Povey, and S.~Khudanpur, ``Librispeech: an asr corpus
  based on public domain audio books,'' in {\em Proc. {IEEE} Intl. Conf. on
  Acoustics, Speech and Signal Processing (ICASSP)}, pp.~5206--5210, IEEE,
  2015.

\end{thebibliography}

\end{document}